\documentclass[prb,twocolumn,superscriptaddress,floatfix,longbibliography]{revtex4}
\usepackage{enumerate}
\usepackage{graphicx}
\usepackage{amsmath}
\usepackage{mathtools}
\usepackage{xcolor}
\usepackage{amsfonts}
\usepackage{amssymb}
\usepackage{units}
\usepackage[normalem]{ulem}
\newcommand{\ket}[1]{| #1 \rangle}

\begin{document}
\pdfoutput=1
\title{Quantum spin systems: toroidal classification and geometric duality}

\author{Vahid Azimi-Mousolou\footnote{Electronic address: vahid.azimi-mousolou@physics.uu.se}}
\affiliation{Department of Applied Mathematics and Computer Science, 
Faculty of Mathematics and Statistics, 
University of Isfahan, Isfahan 81746-73441, Iran}
\affiliation{Department of Physics and Astronomy, Uppsala University, Box 516, 
SE-751 20 Uppsala, Sweden}

\author{Anders Bergman}
\affiliation{Department of Physics and Astronomy, Uppsala University, Box 516, 
SE-751 20 Uppsala, Sweden}

\author{Anna Delin}
\affiliation{Department of Applied Physics, School of Engineering Sciences, 
KTH Royal Institute of Technology, AlbaNova University Center, SE-10691 Stockholm, 
Sweden}
\affiliation{Swedish e-Science Research Center (SeRC), KTH Royal Institute of Technology, 
SE-10044 Stockholm, Sweden}
\affiliation{Wallenberg Initiative Materials Science for Sustainability (WISE), KTH Royal Institute of Technology, SE-10044 Stockholm, Sweden}

\author{Olle Eriksson}
\affiliation{Department of Physics and Astronomy, Uppsala University, Box 516, 
SE-751 20 Uppsala, Sweden}
\affiliation{Wallenberg Initiative Materials Science,
WISE, Uppsala University, Box 516, 
SE-751 20 Uppsala, Sweden}

\author{Manuel Pereiro}
\affiliation{Department of Physics and Astronomy, Uppsala University, Box 516, 
SE-751 20 Uppsala, Sweden}

\author{Danny Thonig}
\affiliation{School of Science and Technology, \"Orebro University, SE-701 82, 
\"Orebro, Sweden}

\author{Erik Sj\"oqvist\footnote{Electronic address: 
erik.sjoqvist@physics.uu.se}}
\affiliation{Department of Physics and Astronomy, Uppsala University, 
Box 516, SE-751 20 Uppsala, Sweden}

\date{\today}% It is always \today, today,

\begin{abstract}
Toroidal classification and geometric duality in quantum spin systems is presented. Through our classification and duality, we reveal that various bipartite quantum features in magnon-systems can manifest equivalently in both bipartite ferromagnetic and antiferromagnetic materials, based upon the availability of relevant Hamiltonian parameters. Additionally, the results highlight the antiferromagnetic regime as an ultra-fast dual counterpart to the ferromagnetic regime, both exhibiting identical capabilities for quantum spintronics and technological applications. Concrete illustrations are provided, demonstrating how splitting and squeezing types of two-mode magnon quantum correlations can be realized across ferro- and antiferromagnetic regimes.
\end{abstract}
\maketitle
\section{Introduction}
Quantum magnonics is a burgeoning interdisciplinary domain merging spintronics, quantum optics, and quantum information science. It explores the quantum properties of magnons, the quanta of spin waves, in magnetic materials, and their interactions with other quantum platforms. By integrating magnons with established quantum systems like cavity photons and superconducting qubits \cite{Lachance-Quirion2019, azimi-mousolou2021, azimi-mousolou2023}, quantum magnonics aims to unlock novel applications in information processing and quantum technologies. Applications range from the development of low-power and high-speed spin-based quantum computing and memory devices to novel approaches for quantum transducers, high-precision measurements, and communications. From fundamental quantum phenomena to practical implementations, quantum magnonics promises to create more efficient and versatile platforms for quantum information science and technology \cite{barman2021, chumak2021, yuan2022}.

So far, quantum magnonics has broadly focused on conventional ordered ferromagnetic (FM) materials \cite{barman2021, chumak2021, yuan2022}. It was only in recent years that some theoretical and experimental advances have shown the comprehensive advantages of antiferromagnetic (AFM) materials over FM in magnonics phenomena \cite{barman2021, chumak2021, yuan2022, jungwirth2016, baltz2018, rezende2019, yuan2020, azimi-mousolou2020, azimi-mousolou2021, azimi-mousolou2023, shiranzaei2023, liu2023}. Abundant room temperature materials, robustness against magnetic fields, ultra-fast THz dynamics due to a much larger exchange field in an antiferromagnet compared to the typical anisotropy field in a ferromagnet, and phenomena such as exchange bias and spin-orbit effects, large and low-energy stabilized magnon squeezing, and entanglement are part of the characteristics of AFM materials. AFM materials have been investigated, especially for their potential in spintronics \cite{jungwirth2016, baltz2018}.

We classify quantum spin systems into toroidal surfaces and demonstrate the existence of a geometric duality between magnonics in FM and AFM materials. We encounter a compelling similarity that bipartite quantum features can emerge equivalently in both bipartite ferromagnetic and antiferromagnetic materials. This equivalence suggests a deeper connection between these seemingly distinct magnetic configurations, opening doors to novel insights and potential applications in the realm of quantum phenomena within magnonics systems.

\section{Bipartite spin model}
\label{Bipartite spin model}
\begin{figure}[h]
\begin{center}
\includegraphics[width=40mm, height=40mm]{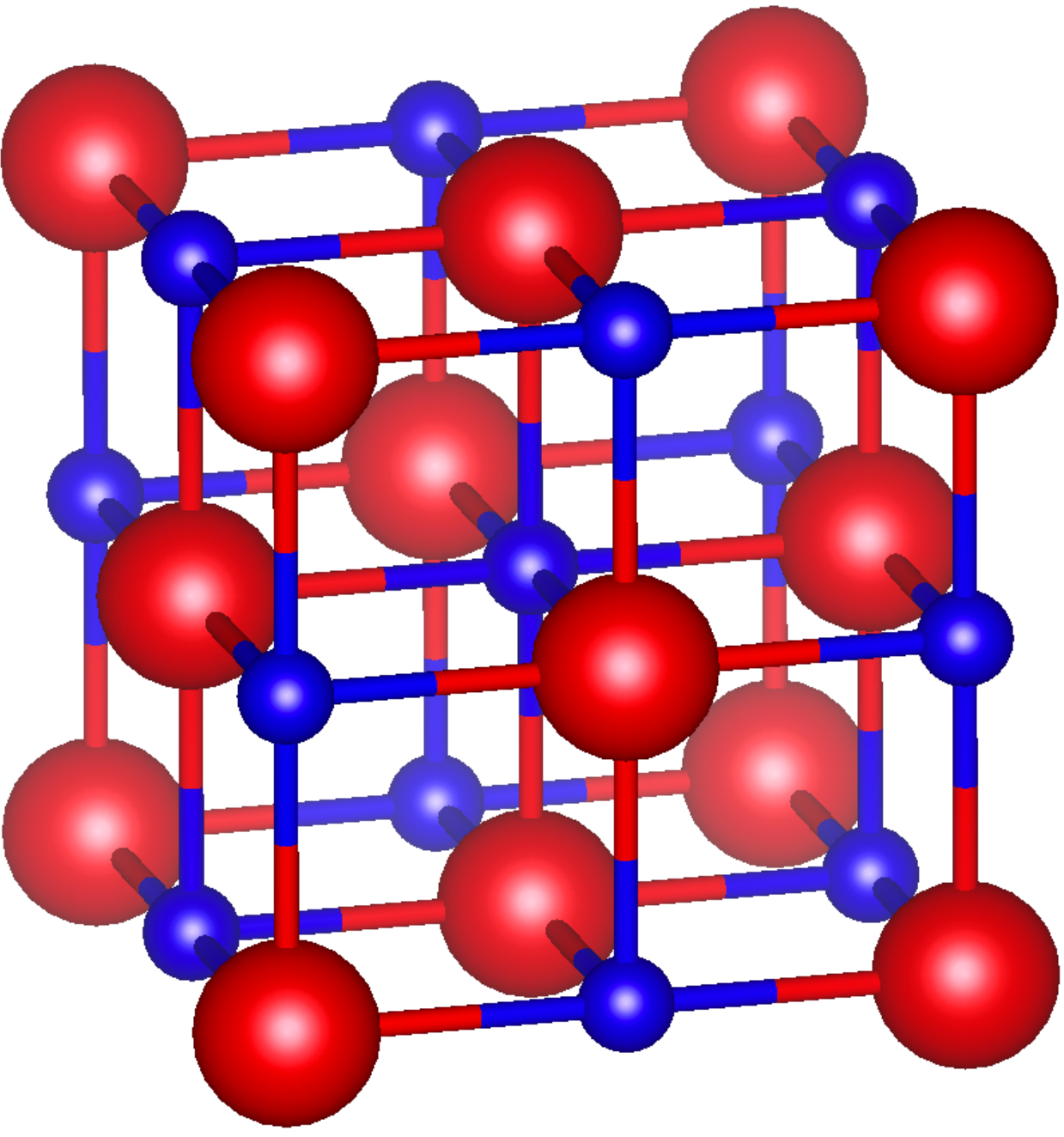}
\end{center}
\caption{(Color online) Schematic illustration of a bipartite spin lattice. Red and blue spheres specify two spin sublattices denoted by $A$ and $B$, respectively, in the present work.}
\label{fig:model}
\end{figure}

We consider a bipartite spin lattice such as the one shown in Fig.\ \ref{fig:model}. 
We assume that the magnetic interactions of the system is described by the spin Hamiltonian
\begin{eqnarray}
H=\sum_{\langle i, j\rangle}\mathbf{S}_{i}\mathbb{I}\mathbf{S}_{j}+\sum_{i}\mathbf{B}\cdot\mathbf{S}_{i}~.
\label{BSM}
\end{eqnarray}
The first term describes the exchange interaction between nearest neighbor spin moments, and the second term describes the coupling to a magnetic field, which, for simplicity, is considered to be applied in the $z$ direction ($\mathbf{B}=B\mathbf{e}_z$). 
For the exchange interaction, we consider an interaction tensor of the form
\begin{eqnarray}
\mathbb{I}=\left(
\begin{array}{ccc}
J+r& K+D& 0   \\
K-D& J-r& 0 \\
0&0& J_z     
\end{array}
\right).\label{eq:FBT}
\end{eqnarray}
In Eq.\eqref{eq:FBT}, the diagonal components are referred to as the Heisenberg exchange with the anisotropic term $r$ describing how the exchange varies in the XY plane. Moreover, the off-diagonal terms are referred to as anisotropic exchange, which can be divided into an antisymmetric term, $\mathbf{D} = D\mathbf{e}_z$, and a symmetric term, $K$, both originating from spin-orbit interaction \cite{szilva2023}. 
For a general spin Hamiltonian, all elements in Eq.~\eqref{eq:FBT} can be non-zero, depending on the symmetry of the system. The resulting magnetic ground state can also be non-collinear depending on the relationship between the antisymmetric and symmetric terms of the Hamiltonian. Here we focus on the fluctuations perpendicular to a starting collinear magnetic state directed in the $\pm\mathbf{e}_z$ direction and thus the corresponding Hamiltonian model shown in Eq.~\eqref{eq:FBT}. In a general bipartite spin-lattice, we denote the spins of the two different sublattices by $\mathbf{S}^{A}_{i}$ with $S_A = |\mathbf{S}^{A}_{i}|$ for each site $i \in A$ (red spheres in Fig.\ \ref{fig:model}) and $\mathbf{S}^{B}_{j}$ with $S_B = |\mathbf{S}^{B}_{j}|$ for each site $j \in B$ (blue spheres in Fig.\ \ref{fig:model}), where $S_A \ne S_B$ in the general ferrimagnetic spin system.

To pursue, we distinguish ferromagnetic (FM) regime by $\mathbb{I}_{ll}< 0$ for $l=1, 2, 3$, and antiferromagnetic (AFM) regime by $\mathbb{I}_{ll}> 0$ for $l=1, 2, 3$. By applying the Holstein-Primakoff transformation, assuming low temperature ($k_BT\ll\min\{|\mathbb{I}_{11}|, |\mathbb{I}_{22}|, |\mathbb{I}_{33}|\}$), the model system can be  described by the effective quadratic magnon Hamiltonian
\begin{eqnarray}
H&=&\sum_{i\in A}\omega_a a_{i}^{\dagger}a_{i} + \sum_{j\in B} \omega_b b_{j}^{\dagger}b_{j}\nonumber\\
&&+\sum_{\langle ij\rangle}\left[\chi a_{i}^{\dagger} b_{j} + \chi^{*}  a_{i} b_{j}^{\dagger}+\Lambda a_{i} b_{j} + \Lambda^{*} a_{i}^{\dagger} b_{j}^{\dagger} \right],\nonumber\\
\label{AFMMH}
\end{eqnarray}
with $\omega_a=\omega+\Delta$,  $\omega_b=\omega-\Delta$ and the star sign representing the complex conjugate. The Hamiltonian parameters in the FM and AFM regimes are given by
\begin{eqnarray}\text{FM:}\ \ \ \ \ \ (\omega, \Delta, \chi, \Lambda)=(B-\tilde{\omega}, \tilde{\Delta}, G, F),\label{FMC}\end{eqnarray}
\begin{eqnarray}\text{AFM:}\ \ \ \ \ \ (\omega, \Delta, \chi, \Lambda)=(\tilde{\omega}, -B-\tilde{\Delta}, F^{*}, G^{*}),\label{AFMC}\end{eqnarray}
where
\begin{eqnarray}
 \tilde{\omega}&=&ZJ_z\left[\frac{S_A+S_B}{2}\right],\ \ \ \ \ G=(J-iD)\sqrt{S_AS_B},\nonumber\\
 \tilde{\Delta}&=&ZJ_z\left[\frac{S_A-S_B}{2}\right],\ \ \ \ \ F=(r-iK)\sqrt{S_AS_B}.\ \ \ \ 
\end{eqnarray}
Here, $Z$ is the coordination number of the spin lattice. By using the Fourier transformations $a_{i}=\sqrt{\frac{2}{N}}\sum_{\mathbf{k}}e^{-i\mathbf{k}\cdot \mathbf{r}_i}a_{\mathbf{k}}$ and $b_{j}=\sqrt{\frac{2}{N}}\sum_{\mathbf{k}}e^{-i\mathbf{k}'\cdot \mathbf{r}_j}b_{\mathbf{k}'}$, where 
$\mathbf{r}_i$ and $\mathbf{r}_j$ denote the position vectors of spins at sites $i$ and $j$ in the lattice, we obtain  
\begin{eqnarray}
H&=& \omega_a a_{\mathbf{k}}^{\dagger}a_{\mathbf{k}} +\omega_b b_{\mathbf{k}}^{\dagger}b_{\mathbf{k}}
\nonumber\\
&&+\chi_{\mathbf{k}}a_{\mathbf{k}}^{\dagger} b_{\mathbf{k}}+\chi^{*}_{\mathbf{k}}a_{\mathbf{k}} b_{\mathbf{k}}^{\dagger}
+\Lambda_{\mathbf{k}}a_{\mathbf{k}} b_{-\mathbf{k}}+\Lambda^{*}_{\mathbf{k}}a_{\mathbf{k}}^{\dagger} b_{-\mathbf{k}}^{\dagger}.\nonumber\\
\label{DMH}
\end{eqnarray}
Here, the ${\mathbf{k}}$ dependence of the parameters is introduced by replacing $G$ and $F$ with $G_{\mathbf{k}}=G\gamma_{\mathbf{k}}$, $F_{\mathbf{k}}=F\gamma_{-\mathbf{k}}$ in the above analysis. The geometric lattice parameter is defined as $\gamma_{\mathbf{k}}=\sum_{\mathbf{\delta}}e^{-i\mathbf{\delta}\cdot \mathbf{k}}$, where the sum is taken over all vectors $\mathbf{\delta}$ connecting each site to the neighbouring sites in the spin lattice. Without loss of generality, we consider the Hamiltonian for a single value of $\mathbf{k}$ in reciprocal $\mathbf{k}$-space, as the state spaces for different values of $\mathbf{k}$ are decoupled from each other. 
The bosonic operators $a_{\mathbf{k}}$, $b_{\mathbf{k}}$ specify localized magnon modes in the corresponding sublattices for a given $\mathbf{k}$.  

\section{Toroidal classification and geometric duality}

We notice that, up to the global phase shifts given by 
\begin{eqnarray}
\left(
\begin{array}{cccc}
  a_{\mathbf{k}}   \\
   b_{\mathbf{k}}      \\
   a^{\dagger}_{-\mathbf{k}}   \\
   b^{\dagger}_{-\mathbf{k}}    
\end{array}
\right)=\left(
\begin{array}{cccc}
1&0&0&0  \\
0& e^{i\nu_{1; \mathbf{k}}}&0& 0  \\
0&0&e^{i\nu_{2; \mathbf{k}}}&0 \\
0&0&0& e^{i\nu_{3; \mathbf{k}}}  
\end{array}
\right)\left(
\begin{array}{cccc}
  \tilde{a}_{\mathbf{k}}   \\
   \tilde{b}_{\mathbf{k}}      \\
   \tilde{a}^{\dagger}_{-\mathbf{k}}   \\
   \tilde{b}^{\dagger}_{-\mathbf{k}}    
\end{array}
\right)
\label{GPS}
\end{eqnarray}
with $\nu_{1; \mathbf{k}}=\arg(\chi_{\mathbf{k}})$, 
$\nu_{2; \mathbf{k}}=\arg(\chi)-\arg(\Lambda)$, and $\nu_{3; \mathbf{k}}=\arg(\Lambda^{*}_{\mathbf{k}})$, which can be adjusted with respect to the phases of $G$, $F$, and $\gamma_{\mathbf{k}}$, the Hamiltonian in Eq. \eqref{DMH},
takes the following form 
\begin{eqnarray}
H&=&\omega_a\tilde{a}_{\mathbf{k}}^{\dagger}\tilde{a}_{\mathbf{k}}+\omega_b\tilde{b}_{\mathbf{k}}^{\dagger}\tilde{b}_{\mathbf{k}}+\tilde{\chi}_{\mathbf{k}}( \tilde{a}_{\mathbf{k}}^{\dagger}\tilde{b}_{\mathbf{k}}+ \tilde{a}_{\mathbf{k}}\tilde{b}_{\mathbf{k}}^{\dagger})\nonumber\\
&&+\tilde{\Lambda}_{\mathbf{k}}( \tilde{a}_{\mathbf{k}}\tilde{b}_{-\mathbf{k}}+ \tilde{a}_{\mathbf{k}}^{\dagger}\tilde{b}_{-\mathbf{k}}^{\dagger})
\label{Htotal}
\end{eqnarray}
with parameters
\begin{eqnarray}\text{FM:}\ \ \ \ \  (\tilde{\chi}_{\mathbf{k}}, \tilde{\Lambda}_{\mathbf{k}})=(|G_{\mathbf{k}}|, |F_{\mathbf{k}}|), \label{FMC1}\end{eqnarray}
\begin{eqnarray} \text{AFM:}\ \ \ \ \  (\tilde{\chi}_{\mathbf{k}}, \tilde{\Lambda}_{\mathbf{k}})=(|F_{\mathbf{k}}|, |G_{\mathbf{k}}|). \label{AFMC1}
\end{eqnarray}
Both Hamiltonians in Eqs. \eqref{DMH} and \eqref{Htotal} describe the same physics, as the magnon modes $(a, b)$ and $(\tilde{a}, \tilde{b})$ represent identical two-mode bosonic systems.

The parameters $\omega_a$ and $\omega_b$ correspond to classical contributions in a sense that in the absence of the interaction parameters $\tilde{\chi}_{\mathbf{k}}$ and $\tilde{\Lambda}_{\mathbf{k}}$, $\omega_a$ and $\omega_b$ represent energy of two separable magnon modes, namely, $a\equiv \tilde{a}$ and $b\equiv \tilde{b}$, each localized in its own sublattice. Once either of $\tilde{\chi}_{\mathbf{k}}$ or $\tilde{\Lambda}_{\mathbf{k}}$ is turned on (which is the case for all physical magnetic systems) the localized magnon modes become hybridized and give rise to nonlocal and entangled magnon modes. Put succinctly, $\tilde{\chi}_{\mathbf{k}}$ and $\tilde{\Lambda}_{\mathbf{k}}$ characterize significant quantum features in the Hamiltonian in Eq.\ \eqref{Htotal}. However, one should point out that they give rise to different quantum features. While $\tilde{\chi}_{\mathbf{k}}$ correspond to magnon splitting in the system and derive discrete variable 
two-mode magnon entanglement between the localized magnon modes $\tilde{a}_{\mathbf{k}}$ and $\tilde{b}_{\mathbf{k}}$ with parallel momenta, $\tilde{\Lambda}_{\mathbf{k}}$ is responsible for two-mode magnon squeezing and continuous variable two-mode magnon entanglement between the localized magnon modes $\tilde{a}_{\mathbf{k}}$ and $\tilde{b}_{-\mathbf{k}}$ with antiparallel momenta. Recently, in a series of papers \cite{azimi-mousolou2020, azimi-mousolou2021, azimi-mousolou2023, liu2023, shiranzaei2023}, we have analyzed fundamental and practical aspects of two-mode quantum squeezing and continuous variable entanglement in antiferromagnetic materials, with a particular focus on designing setups for experimental verification in the lab.

An important and intriguing observation from the preceding analysis is that, irrespective of the type of entanglement or nonlocality measures, quantum correlation delineates a clear foliation of the manifold of coupling parameters specified by $(J, D, r, K)$ into two-dimensional compact torus leaves in both FM and AFM materials. For example, in the case of FM materials, all bipartite materials with a given pair of localized magnon dispersions $\omega_a$ and $\omega_b$ contribute an identical amount of quantum correlations between the two magnon modes $a$ and $b$ at each point of the Brillouin zone, corresponding to the fixed value of $|\gamma_{\mathbf{k}}|$, if and only if the variable parameters $J$, $D$, $r$, and $K$ satisfy the following toric equations
\begin{eqnarray}
J^2 + D^2 &=& R_{1}^{2}, \nonumber\\
r^2 + K^2 &=& R_{2}^{2},
\label{FMTC}
\end{eqnarray}
for fixed radii $R_{1}$ and $R_{2}$ such that $|G|=R_{1}\sqrt{S_AS_B}$ 
and $|F|=R_{2}\sqrt{S_AS_B}$.
It follows from Eq. \eqref{FMC1} that all parameters satisfying Eq. \eqref{FMTC} specify the same two-mode bosonic Hamiltonian in Eq. \eqref{Htotal}. The criterion in Eq. \eqref{FMTC} identifies a two-dimensional torus with radii $R_{1}$ and $R_{2}$ in the four-dimensional parameter space.
This demonstrates that quantum correlation classifies bipartite magnonics systems in the ferromagnetic regime into equivalent toric classes.

Similarly, in the antiferromagnetic regime, we observe classifications of bipartite magnonic systems into topological toric equivalent classes. 
By following Eq. \eqref{AFMC1}, the criteria equivalent to Eq. \eqref{FMTC} in the AFM case are given by
\begin{eqnarray}
r^2 + K^2 &=& R_{1}^{2}, \nonumber\\
J^2 + D^2 &=& R_{2}^{2},
\label{AFMTC}
\end{eqnarray}
such that $|F|=R_{1}\sqrt{S_AS_B}$ and $|G|=R_{2}\sqrt{S_AS_B}$. This indicate the variable parameters $J$, $D$, $r$, and $K$ in the AFM regime, which satisfy the toric equations in Eq. \eqref{AFMTC}, give rise to the same two-mode bosonic Hamiltonian in Eq. \eqref{Htotal} and thus contribute an identical amount of quantum correlations between the two magnon modes $a$ and $b$, with localized magnon dispersions $\omega_a$ and $\omega_b$, at each point of the Brillouin zone corresponding to the fixed values of $|\gamma_{\mathbf{k}}|$.

\begin{figure}[h]
\begin{center}
\includegraphics[width=85mm, height=43mm]{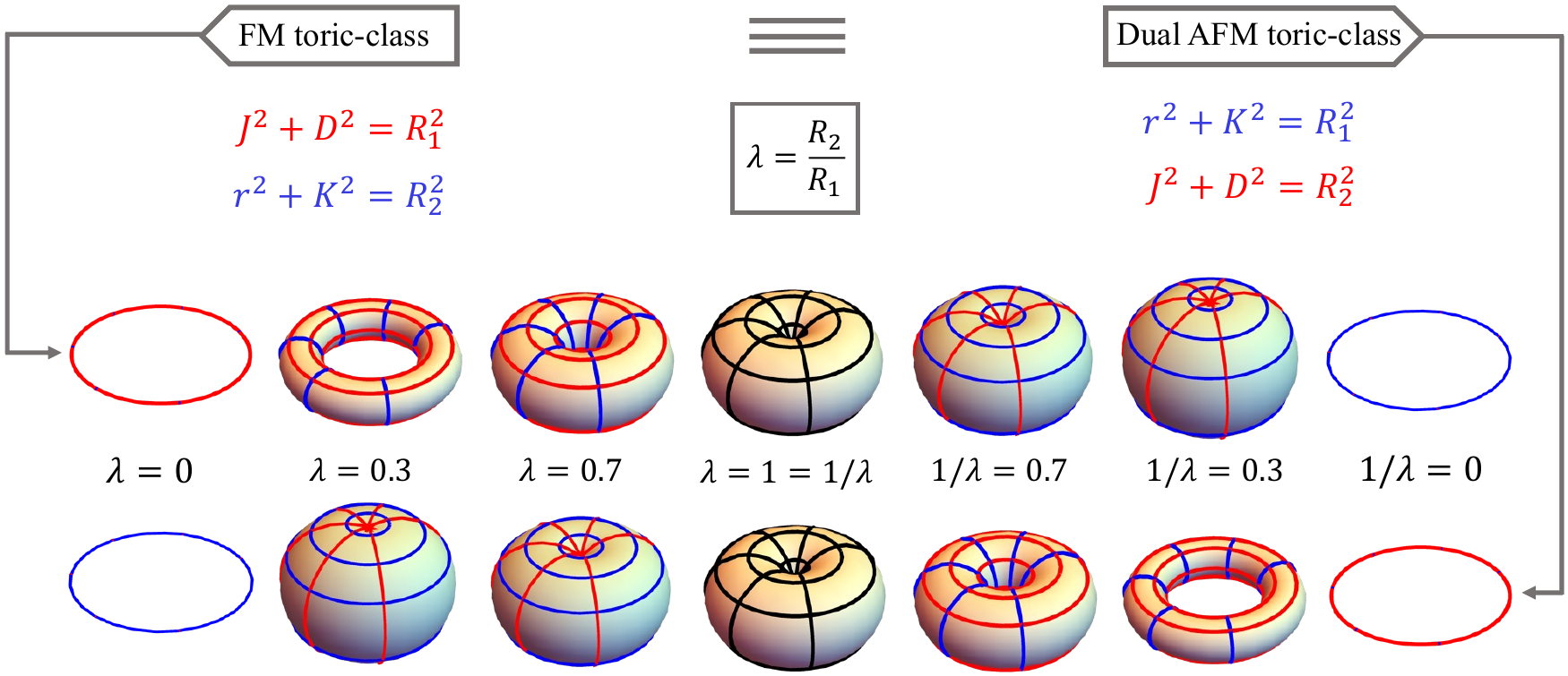}
\end{center}
\caption{(Color online) Schematic illustration of the toroidal classification of quantum spin systems and quantum geometric duality in quantum magnonics. For each FM toric class in the upper row, there exists its dual AFM toric class in the lower row, which is equivalent in the sense that both identify the same bipartite magnonics quantum correlation in their respective regimes. The longitude and meridian circles, and consequently the roles of the principal curvatures, are switched from FM toric classes to AFM toric classes. However, both toric classes in each dual pair share the same Gauss curvature and mean curvature, confirming the global geometric characteristics of quantum correlations.}
\label{fig:TDM}
\end{figure}

The toric characteristic equations in Eqs. \eqref{FMTC} and \eqref{AFMTC} reveal the presence of a quantum geometric duality between bipartite ferro- and antiferromagnetic materials. Specifically, for any given ferromagnetic material with a bipartite magnonics structure, there exists its dual antiferromagnetic counterpart that gives rise to the same quantum correlations between the two magnons in the system, and vice versa. This follows from the fact that the two sets of toric equations in Eqs. \eqref{FMTC} and \eqref{AFMTC}, within their respective parameter regimes, identify the same quantum-correlated two-mode bosonic Hamiltonian for a given pair of magnon dispersions $\omega_a$ and $\omega_b$, and the geometric lattice parameter $|\gamma_{\mathbf{k}}|$. We note that the longitude and meridian circles in Eq. \eqref{AFMTC}, compared to Eq. \eqref{FMTC}, are interchanged. This means that the radii, and thus the principal curvatures, of an AFM toric class are swapped compared to its dual FM toric class. However, both FM and AFM dual toric classes have the same Gauss curvature and mean curvature. Such consistency is reflected in the two-mode magnon quantum correlations that both dual classes identify, confirming the global geometric characteristics of quantum correlations.
Fig.\ \ref{fig:TDM} illustrates the toroidal classification and geometric duality.

We end this section by noting that most magnetic materials, FM or AFM, are well described by the spin Hamiltonian outlined in Eqs.\ \eqref{BSM} and \eqref{eq:FBT}, and that in general all parameters of Eq.\ \eqref{eq:FBT} are non-zero. Furthermore, the parameters $D$ and $K$ are coupled to the spin-orbit interaction and are therefore of relativistic quantum origin \cite{szilva2023}. This implies that the analysis presented here is relevant for a large group of magnetic systems. Comparing the typical strength of the interactions of Eq.\ \eqref{eq:FBT}, $J$ normally dominates, while $D$ and $K$ can be of order 10 \% of $J$, and $r$ is normally the smallest interaction.

\section{Examples}

To better clarify the quantum correlation-induced geometric duality in magnonics systems, we consider two specific cases that have recently been the focus of quantum magnonics research \cite{azimi-mousolou2021, azimi-mousolou2023, barman2021, chumak2021, yuan2022, azimi-mousolou2020, shiranzaei2023, liu2023, kamra2020}.

{\it Splitting quantum correlation:} Let us focus only on the splitting type of couplings and analyze the type of magnon entanglement they give rise to. In this case the Hamiltonian reads,
\begin{eqnarray}
H^{\text{SP}}=\omega_a\tilde{a}_{\mathbf{k}}^{\dagger}\tilde{a}_{\mathbf{k}}+\omega_b\tilde{b}_{\mathbf{k}}^{\dagger}\tilde{b}_{\mathbf{k}}+\tilde{\chi}_{\mathbf{k}}( \tilde{a}_{\mathbf{k}}^{\dagger}\tilde{b}_{\mathbf{k}}+ \tilde{a}_{\mathbf{k}}\tilde{b}_{\mathbf{k}}^{\dagger}),
\label{HtotalSP}
\end{eqnarray}
where $\tilde{\chi}_{\mathbf{k}}=|G_{\mathbf{k}}|=|\gamma_{\mathbf{k}}|\sqrt{J^{2}+ D^{2}}$ for FM interactions, and $\tilde{\chi}_{\mathbf{k}}=|F_{\mathbf{k}}|=|\gamma_{\mathbf{k}}|\sqrt{r^{2}+ K^{2}}$ for AFM interactions.
The Hamiltonian $H^{\text{SP}}$ can be diagonalized through the $SU(2)$ transformation 
\begin{eqnarray}
\left(
\begin{array}{cc}
  \tilde{a}_{\mathbf{k}}   \\
   \tilde{b}_{\mathbf{k}}    
\end{array}
\right)=\left(
\begin{array}{cc}
  u_{\mathbf{k}}& v_{\mathbf{k}}   \\
-v_{\mathbf{k}}& u_{\mathbf{k}}       
\end{array}
\right)\left(
\begin{array}{cc}
  \alpha_{\mathbf{k}}   \\
   \beta_{\mathbf{k}}     
\end{array}
\right),
\label{eq:UT}
\end{eqnarray}
with $u_{\mathbf{k}}=\cos\theta_{\mathbf{k}}$ and $v_{\mathbf{k}}=\sin\theta_{\mathbf{k}}$. By applying this transformation, we obtain the diagonal Hamiltonian
\begin{eqnarray}
H^{\text{SP}}=\omega_{\alpha_{\mathbf{k}}}\alpha_{\mathbf{k}}^{\dagger}\alpha_{\mathbf{k}}+\omega_{\beta_{\mathbf{k}}}\beta_{\mathbf{k}}^{\dagger}\beta_{\mathbf{k}},
\label{Htotal1}
\end{eqnarray}
in the hybridized magnon modes $\alpha_{\mathbf{k}}$ and $\beta_{\mathbf{k}}$, where 
\begin{eqnarray}
\omega_{\alpha_{\mathbf{k}}}&=&\omega+\Delta\cos2\theta_{\mathbf{k}}-\tilde{\chi}_{\mathbf{k}}\sin 2\theta_{\mathbf{k}}\nonumber\\
\omega_{\beta_{\mathbf{k}}}&=&\omega-\Delta\cos2\theta_{\mathbf{k}}+\tilde{\chi}_{\mathbf{k}}\sin 2\theta_{\mathbf{k}}
\end{eqnarray}
with $\omega$ and $\Delta$ given by Eqs.\ \eqref{FMC} and \eqref{AFMC}.
Here $\theta_{\mathbf{k}}=\pm\pi/4$ for $\Delta=0$, and 
\begin{eqnarray}
\tan\theta_{\mathbf{k}}=\frac{1-\sqrt{1+|\Gamma_{\mathbf{k}}|^{2}}}{\Gamma_{\mathbf{k}}},\ \ \ \ \ \ \ \ \Gamma_{\mathbf{k}}=\frac{\tilde{\chi}_{\mathbf{k}}}{\Delta}
\label{SPP}
\end{eqnarray}
for $\Delta\ne 0$ .

From Eq. \eqref{Htotal1}, we obtain the energy eigenstates
\begin{eqnarray}
\ket{\psi_{mn}^{\text{SP}}}&=&\ket{m; \alpha_{\mathbf{k}}}\ket{n;\beta_{\mathbf{k}}}\nonumber\\
&=&\frac{1}{\sqrt{m!n!}}[\beta_{\mathbf{k}}^{\dagger}]^{n}[\alpha_{\mathbf{k}}^{\dagger}]^{m}\ket{0; \alpha_{\mathbf{k}}}\ket{0; \beta_{\mathbf{k}}}
\nonumber\\
\end{eqnarray}
in the $(\alpha, \beta)$ modes, where $\ket{0; \alpha_{\mathbf{k}}}$ and $\ket{0; \beta_{\mathbf{k}}}$ are vacuum states of the bosonic operators 
$\alpha_{\mathbf{k}}$ and $\beta_{\mathbf{k}}$, respectively. These energy eigenstates take the following form 

\begin{eqnarray}
\ket{\psi^{\text{SP}}_{mn}}&=&\sum_{p=0}^{m}c_{mn}^{(\mathbf{k}; 1)}(p)\ket{m-p; \tilde{a}_{\mathbf{k}}}\ket{n+p; \tilde{b}_{\mathbf{k}}}+
\nonumber\\&&\sum_{q=1}^{n}c_{mn}^{(\mathbf{k}; 2)}(q)\ket{m+q; \tilde{a}_{\mathbf{k}}}\ket{n-q; \tilde{b}_{\mathbf{k}}}
\label{EESSP}
\end{eqnarray}
 in the $(\tilde{a}, \tilde{b})$ modes, where we obtain
\begin{eqnarray}
c_{mn}^{(\mathbf{k}; 1)}(p)&=&\frac{1}{\sqrt{m!n!}}\sum_{l=0}^{\min\{m-p, n\}}c^{\mathbf{k}}_{mn}(p+l, l),\nonumber\\
c_{mn}^{(\mathbf{k}; 2)}(q)&=&\frac{1}{\sqrt{m!n!}}\sum_{l=0}^{\min\{m, n-q\}}c^{\mathbf{k}}_{mn}(l, q+l)
\end{eqnarray}
with 
\begin{eqnarray}
c^{\mathbf{k}}_{mn}(x, y)&=&(-1)^{x}\left(
\begin{array}{cc}
  m   \\
  x   
\end{array}
\right)\left(
\begin{array}{cc}
  n  \\
  y    
\end{array}
\right)v_{\mathbf{k}}^{x+y}u_{\mathbf{k}}^{m+n-(x+y)}\nonumber\\
&&\times\sqrt{(m-x+y)!(n-y+x)!}.
\end{eqnarray}
From these expressions, it is obvious that all the weight coefficients $c_{mn}^{(\mathbf{k}; 1)}(p)$ and $c_{mn}^{(\mathbf{k}; 2)}(q)$ of the energy eigenstates in Eq. \eqref{EESSP} only depend on the splitting parameter $\theta_{\mathbf{k}}$ introduced in Eq. \eqref{SPP}. Consequently, the quantum correlations present in these states, measured through, for instance, the entanglement entropy
\begin{eqnarray}
E^{\text{SP}}_{mn}(\theta_{\mathbf{k}})&=&-\sum_{p=0}^{m}|c_{mn}^{(\mathbf{k}; 1)}(p)|^{2}\log |c_{mn}^{(\mathbf{k}; 1)}(p)|^{2}
\nonumber\\&&-\sum_{q=1}^{n}|c_{mn}^{(\mathbf{k}; 2)}(q)|^{2}\log |c_{mn}^{(\mathbf{k}; 2)}(q)|^{2}
\end{eqnarray}
also depend solely on the splitting parameter $\theta_{\mathbf{k}}$. Note that since $(\tilde{a}, \tilde{b})$ modes are equivalent to $(a, b)$ modes up to phase shifts given by Eq.\ \eqref{GPS}, the entanglement entropy in both pairs of modes is exactly the same.

The splitting aspect of the parameter $\theta_{\mathbf{k}}$ is related to the splitting nature of the coupling parameter $\tilde{\chi}_{\mathbf{k}}$ through Eq.\ \eqref{SPP}, which leads to the eigenenergy of the splitting Hamiltonian being a finite linear combination of magnon number states in the localized magnon modes $(\tilde{a}, \tilde{b})$, or equivalently $(a, b)$. Finite superpositions are naturally relevant for discrete variable quantum computation. Our analysis here shows that both ferromagnetic and antiferromagnetic materials allow for the same splitting quantum correlation between two magnon modes. However, in the two regimes, this quantum correlation is derived by different parameters. While in the FM regime, $\theta_{\mathbf{k}}$ and thus the splitting quantum correlation are given by $J$, $D$, and $\Delta=ZJ_z\left[\frac{S_A-S_B}{2}\right]$, in the dual AFM regime, $\theta_{\mathbf{k}}$ and thus the splitting quantum correlation are given by $r$, $K$, and $\Delta=B-ZJ_z\left[\frac{S_A-S_B}{2}\right]$. The splitting quantum correlation exemplifies the aforementioned quantum geometric duality, explicitly the dual red and blue circles in the first column from the left in Fig.\ \ref{fig:TDM}, characterized by the radii $R_1=\sqrt{J^{2}+ D^{2}}=\sqrt{r^{2}+ K^{2}}\ne 0$ and $R_2= 0$.

{\it Squeezing quantum correlation:} Let us consider the squeezing type of coupling described by the Hamiltonian
\begin{eqnarray}
H^{\text{SQ}}=\omega_a\tilde{a}_{\mathbf{k}}^{\dagger}\tilde{a}_{\mathbf{k}}+\omega_b\tilde{b}_{\mathbf{k}}^{\dagger}\tilde{b}_{\mathbf{k}}+\tilde{\Lambda}_{\mathbf{k}}(\tilde{a}_{\mathbf{k}}\tilde{b}_{-\mathbf{k}}+\tilde{a}_{\mathbf{k}}^{\dagger}\tilde{b}_{-\mathbf{k}}^{\dagger}),\ \ \ 
\label{HtotalSQ}
\end{eqnarray}
where $\tilde{\Lambda}_{\mathbf{k}}=|F_{\mathbf{k}}|=|\gamma_{\mathbf{k}}|\sqrt{r^{2}+ K^{2}}$ for FM interactions, and $\tilde{\Lambda}_{\mathbf{k}}=|G_{\mathbf{k}}|=|\gamma_{\mathbf{k}}|\sqrt{J^{2}+ D^{2}}$  for AFM interactions.

Under $SU(1,1)$ Bogoliubov transformation
 \begin{eqnarray}
\left(
\begin{array}{cc}
  \tilde{a}_{\mathbf{k}}    \\
   \tilde{b}_{-\mathbf{k}}^{\dagger}       
\end{array}
\right)=\left(
\begin{array}{cc}
\eta_{\mathbf{k}}& \zeta_{\mathbf{k}}    \\
\bar{\zeta}_{\mathbf{k}}& \bar{\eta}_{\mathbf{k}}      
\end{array}
\right)\left(
\begin{array}{cc}
  \tilde{\alpha}_{\mathbf{k}}    \\
   \tilde{\beta}_{-\mathbf{k}}^{\dagger}       
\end{array}
\right),
\label{eq:FBTb}
\end{eqnarray}
where $\eta_{\mathbf{k}} =\cosh(r_{\mathbf{k}})$ and $\zeta_{\mathbf{k}} = \sinh(r_{\mathbf{k}})$ with
\begin{eqnarray}
\tanh r_{\mathbf{k}}&=&\frac{1-\sqrt{1-|\tilde{\Gamma}_{\mathbf{k}}|^{2}}}{\tilde{\Gamma}_{\mathbf{k}}},\ \ \ 
\tilde{\Gamma}_{\mathbf{k}}=\frac{\tilde{\Lambda}_{\mathbf{k}}}{\omega},
\label{SQPP}
\end{eqnarray}
we obtain the following diagonal form of the Hamiltonian
 \begin{eqnarray}
H^{\text{SQ}} = 
\omega_{\tilde{\alpha}_{\mathbf{k}}}\tilde{\alpha}_{\mathbf{k}}^{\dagger}\tilde{\alpha}_{\mathbf{k}} +
\omega_{\tilde{\beta}_{-\mathbf{k}}}\tilde{\beta}_{-\mathbf{k}}^{\dagger} \tilde{\beta}_{-\mathbf{k}},
\label{DHH}
\end{eqnarray}
provided $|\tilde{\Gamma}_{\mathbf{k}}|<1$. The magnon dispersion relations are  
\begin{eqnarray}
\omega_{\tilde{\alpha}_{\mathbf{k}}}&=&\omega\cosh(2 r_{\mathbf{k}})+\tilde{\Lambda}_{\mathbf{k}}\sinh(2 r_{\mathbf{k}})+\Delta\nonumber\\
\omega_{\tilde{\beta}_{-\mathbf{k}}}&=&\omega\cosh(2 r_{\mathbf{k}})+\tilde{\Lambda}_{\mathbf{k}}\sinh(2 r_{\mathbf{k}})-\Delta,\ \ \ \ \ \ \ \ 
\end{eqnarray}
where $\omega$ and $\Delta$ are given in Eqs.\ \eqref{FMC} and \eqref{AFMC}.

Eq. \eqref{HtotalSQ} implies that the energy eigenstates take the following form 
\begin{eqnarray}
\ket{\psi_{mn}^{\text{SQ}}}&=&\ket{m; \tilde{\alpha}_{\mathbf{k}}}\ket{n;\tilde{\beta}_{-\mathbf{k}}}\nonumber\\
&=&\frac{1}{\sqrt{m!n!}}[\tilde{\beta}_{\mathbf{k}}^{\dagger}]^{n}[\tilde{\alpha}_{\mathbf{k}}^{\dagger}]^{m}\ket{0; \tilde{\alpha}_{\mathbf{k}}}\ket{0; \tilde{\beta}_{-\mathbf{k}}}
\label{SQEES}
\end{eqnarray}
in the $(\tilde{\alpha}, \tilde{\beta})$ modes, where $\ket{0; \tilde{\alpha}_{\mathbf{k}}}$ and $\ket{0;\tilde{\beta}_{-\mathbf{k}}}$ are vacuum states of the bosonic operators 
$\tilde{\alpha}_{\mathbf{k}}$ and $\tilde{\beta}_{-\mathbf{k}}$ respectively. In the localized  $(\tilde{a}, \tilde{b})$ magnon modes, these energy eigenstates take the following form 

\begin{eqnarray}
&&\ket{\psi_{mn}^{\text{SQ}}}=
\sum_{p=0}^{\infty} \tilde{c}_{mn}^{\mathbf{k}}(p)\ket{p+\delta l; \tilde{a}_{\mathbf{k}}} 
\ket{p; \tilde{b}_{-\mathbf{k}}},\ \ \ \ \ \ m\ge n
\nonumber\\
&&\ket{\psi_{mn}^{\text{SQ}}}=
\sum_{p=0}^{\infty} \tilde{c}_{mn}^{\mathbf{k}}(p)\ket{p; \tilde{a}_{\mathbf{k}}} 
\ket{p+\delta l; \tilde{b}_{-\mathbf{k}}},\ \ \ \ \ \ m\le n
\nonumber\\
\label{EES}
\end{eqnarray}
with $\delta l=|m-n|$ and the probability amplitudes
\begin{eqnarray}
\tilde{c}_{mn}^{\mathbf{k}}(p)=\frac{1}{\sqrt{m!n!}}
\left(\frac{1}{\eta_{\mathbf{k}}}\right)^{\delta l}\left(\frac{1}{\eta_{\mathbf{k}}\zeta_{\mathbf{k}}}\right)^{l}q^{(l, \delta l)}_{p; \mathbf{k}}\tilde{c}_{00}^{\mathbf{k}}(p),\nonumber\\
\end{eqnarray}
where $l=\min\{m, n\}$,
$\tilde{c}_{00}^{\mathbf{k}}(p)=\frac{\tanh^{p}(r_{\mathbf{k}}}{\cosh(r_{\mathbf{k}})})$,
 and $q^{(l, \delta l)}_{p; \mathbf{k}}$ satisfies the following recursive relations 
\begin{eqnarray}
q^{(l, \delta l)}_{p; \mathbf{k}}&=&\eta_{\mathbf{k}}^{2}\sqrt{p+\delta l}q^{(l, \delta l-1)}_{p; \mathbf{k}}
-\zeta_{\mathbf{k}}^{2}\sqrt{p+1}q^{(l, \delta l-1)}_{p+1; \mathbf{k}}\nonumber\\
q^{(l, 0)}_{p; \mathbf{k}}&=&p\eta_{\mathbf{k}}^{4}q^{(l-1, 0)}_{p-1; \mathbf{k}}-(2p+1)\eta_{\mathbf{k}}^{2}\zeta_{\mathbf{k}}^{2}q^{(l-1,0)}_{p; \mathbf{k}}\nonumber\\
&&+(p+1)\zeta_{\mathbf{k}}^{4}q^{(l-1,0)}_{p+1; \mathbf{k}},
\end{eqnarray}
with $q^{(0, 0)}_{p; \mathbf{k}}=1$ for each $p$ as the initial value condition.

All the weight coefficients $\tilde{c}_{mn}^{\mathbf{k}}(p)$ appearing in Eq. \eqref{SQEES} are solely dependent on the squeezing parameter $r_{\mathbf{k}}$ introduced in Eq.\ \eqref{SQPP}. This implies that the quantum correlations within these states, quantified by measures such as the entanglement entropy,
\begin{eqnarray}
E^{\text{SQ}}_{mn}(r_{\mathbf{k}})=-\sum_{p=0}^{\infty}| \tilde{c}_{mn}^{\mathbf{k}}(p)|^{2}\log | \tilde{c}_{mn}^{\mathbf{k}}(p)|^{2},
\end{eqnarray}
also rely entirely on squeezing parameter $r_{\mathbf{k}}$. Since the modes $(\tilde{a}, \tilde{b})$ are equivalent to $(a, b)$ modes, up to phase shifts as given by Eq. \eqref{GPS}, the entanglement entropy for both pairs of modes remains identical

The squeezing aspect of the parameter $r_{\mathbf{k}}$ is associated with the squeezing nature of the coupling parameter $\tilde{\Lambda}_{\mathbf{k}}$ via Eq. \eqref{SQPP}. This squeezing nature results in the eigenenergy of the Hamiltonian $H^{\text{SQ}}$ being an infinite linear combination of magnon number states in the localized magnon modes $(\tilde{a}, \tilde{b})$, or equivalently $(a, b)$. Infinite superpositions are naturally relevant for continuous variable quantum computation. Our analysis demonstrates that both ferromagnetic and antiferromagnetic materials exhibit the same squeezing quantum correlation between two magnon modes. However, this quantum correlation is determined by different parameters for the two systems. In the FM regime, $r_{\mathbf{k}}$ and thus the squeezing quantum correlation are governed by $r$, $K$, and $\omega=B-ZJ_z\left[\frac{S_A+S_B}{2}\right]$, whereas in the dual AFM regime, $r_{\mathbf{k}}$ and thus the squeezing quantum correlation are determined by $J$, $D$, and $\omega=ZJ_z\left[\frac{S_A+S_B}{2}\right]$. 
The squeezing quantum correlation exemplifies the aforementioned quantum geometric duality, explicitly the dual blue and red circles in the first column from the right in Fig.\ \ref{fig:TDM}, characterized by the radii $R_1=0$ and $R_2=\sqrt{r^{2}+ K^{2}}=\sqrt{J^{2}+ D^{2}}\ne 0$.

\section{conclusion}
In summary, our study delineates a toroidal classification of quantum spin systems and elucidates the geometric duality inherent in bipartite spin systems within the realm of quantum magnonics. Each pair of dual toric classes is specified by a unique pair of principal curvatures, where the roles of the principal curvatures are swapped between a toric class and its dual. However, both dual toric classes identify the same Gauss curvature, mean curvature, and two-mode magnon quantum correlations. Such consistency confirms the global geometric characteristics of quantum correlations in bipartite spin systems. This duality underscores the parity of various bipartite quantum characteristics across both ferromagnetic and antiferromagnetic materials possessing a bipartite structure, contingent upon the availability and appropriateness of the respective Hamiltonian parameters. Furthermore, our findings emphasize the antiferromagnetic domain as an ultra-fast dual counterpart to the ferromagnetic domain, showcasing equivalent potential for quantum spintronic and technological advancements. To gain insights into how the geometric duality manifests in various quantum spin systems and understand its implications, we examine concrete examples that illustrate the emergence of splitting and squeezing patterns of two-mode magnon quantum correlations across both ferromagnetic and antiferromagnetic systems.

The classification and geometric duality demonstrated here clarify the availability of a broader range of materials for advances in quantum spintronics and quantum magnonics, as well as their applications in quantum information processing. This study is useful for categorizing and gaining insights into the physical behaviors and properties of various quantum spin systems, identifying new materials with specific and desirable quantum properties, predicting and discovering new quantum phenomena with technological implications, engineering quantum states with desired properties, and deepening our understanding of quantum entanglement and geometric properties, particularly in quantum spin systems. Therefore, experimental verification of such classification and duality would offer valuable opportunities to further progress in these fields and make new discoveries. The experimental setups proposed and designed in \cite{azimi-mousolou2021, azimi-mousolou2023, Lachance-Quirion2019, Lachance-Quirion2020} could facilitate this experimental verification.

\section*{Acknowledgments}
Financial support from the
Swedish Research Council (Vetenskapsr{\aa}det, VR) Grant No. 2016-05980, Grant No. 2019-05304, Grant No. 2019-03666, Grant No. 2023-04899, and Grant No. 2023-04239, and the Knut and Alice Wallenberg foundation Grant No. 2018.0060, Grant No. 2021.0246, and Grant No. 2022.0108 is acknowledged.  
The Wallenberg Initiative Materials Science for Sustainability (WISE) funded by the Knut and Alice Wallenberg Foundation is also acknowledged.
O.E. also acknowledges The European Research Council (ERC), eSSENCE and STandUPP.


\begin{thebibliography}{} 

\bibitem{Lachance-Quirion2019} D. Lachance-Quirion, Y. Tabuchi, A. Gloppe, K. Usami, and Y. Nakamura, Hybrid quantum systems based on magnonics, Appl. Phys. Express {\bf 12}(7), 070101 (2019).
\bibitem{azimi-mousolou2021} V. Azimi-Mousolou, Y. Liu, A. Bergman, A. Delin, O. Eriksson, M. Pereiro, D. Thonig, and E. Sj\"oqvist, Magnon-magnon entanglement and its quantification via a microwave cavity,
Phys. Rev. B {\bf 104}, 224302 (2021).
\bibitem{azimi-mousolou2023} V. Azimi-Mousolou, A. Bergman, A. Delin, O. Eriksson, M. Pereiro, D. Thonig, and E. Sj\"oqvist, Transmon probe for quantum characteristics of magnons in antiferromagnets,
Phys. Rev. B {\bf 108}, 094430 (2023).
\bibitem{barman2021} A. Barman et al., The 2021 Magnonics Roadmap, J. Phys.: Condens. Matter {\bf 33}, 413001 (2021); 
\bibitem{chumak2021} A. V. Chumak et al., Roadmap on Spin-Wave Computing concepts, IEEE Trans. Quant. Eng. {\bf 2}, 1–10 (2021).
\bibitem{yuan2022} [4] H. Y. Yuan, Y. Cao, A. Kamra, R. A. Duine, P. Yan, Quantum magnonics: When magnon spintronics meets quantum information science, Phys. Rep. {\bf 965}, 1 (2022). 
\bibitem{jungwirth2016} T. Jungwirth, X. Marti, P. Wadley, and J. Wunderlich, Antiferromagnetic spintronics, Nat. Nanotech. {\bf 11}, 231–241 (2016). 
\bibitem{baltz2018} V. Baltz, A. Manchon, M. Tsoi, T. Moriyama, T. Ono, and Y. Tserkovnya, Antiferromagnetic spintronics, Rev. Mod. Phys. {\bf 90}, 015005 (2018).
\bibitem{rezende2019} S. M. Rezende, A. Azevedo, and R. L. Rodr\'iguez-Su\'arez,  Introduction to antiferromagnetic magnons, J. Appl. Phys. {\bf 126}, 151101 (2019). 
\bibitem{yuan2020} H. Y. Yuan, Z. Yuan, R. A. Duine, X. R. Wang, Recent progress in antiferromagnetic dynamics, EPL {\bf 132}, 57001 (2020).
\bibitem{azimi-mousolou2020} V. Azimi-Mousolou, A. Bagrov, A. Bergman, A. Delin, O. Eriksson, Y. Liu, M. Pereiro, D. Thonig, and E. Sj\"oqvist, Hierarchy of magnon mode entanglement in antiferromagnets, Phys. Rev. B {\bf 102}, 224418 (2020).
\bibitem{shiranzaei2023} M. Shiranzaei, J. Fransson, and V. Azimi-Mousolou, Temperature-anisotropy conjugate magnon squeezing in antiferromagnets, Phys. Rev. B {\bf 108}, 144302 (2023).
\bibitem{liu2023} Y. Liu, A. Bergman, A. Bagrov, A. Delin, D. Thonig, M. Pereiro, O. Eriksson, S. Streib, E. Sj\"oqvist, and V. Azimi-Mousolou, Tunable phonon-driven magnon–magnon entanglement at room temperature, New J. Phys. {\bf 25}, 113032 (2023).
\bibitem{szilva2023} A. Szilva, Y. Kvashnin, E. A. Stepanov, L. Nordstr\"om, O. Eriksson, A. I. Lichtenstein, and M. I. Katsnelson, Quantitative theory of magnetic interactions in solids, Rev. Mod. Phys. {\bf 95}, 035004 (2023).
\bibitem{kamra2020} A. Kamra, B. Wolfgang, and B. Arne, Magnon-squeezing as a niche of quantum magnonics, Appl. Phys. Lett. {\bf 117}, 090501 (2020).
\bibitem{Lachance-Quirion2020} D. Lachance-Quirion, S. P. Wolski, Y. Tabuchi, S. Kono, K. Usami, Y. Nakamura, Entanglement-based single-shot detection of a single magnon with a superconducting qubit, Science {\bf 367} (6476), 425–428 (2020).
\end{thebibliography}
\end{document}